# Spin-Diffusion Lengths in Dilute Cu(Ge) and Ag(Sn) Alloys.

Q. Fowler,* B. Richard,* A. Sharma, N. Theodoropoulou, R. Loloee, W.P. Pratt Jr., and J. Bass
Department of Physics and Astronomy, Michigan State University
East Lansing, MI 48824
\* These two authors made comparable contributions to this research.

We use current-perpendicular-to-plane (CPP) exchange-biased spin-valves to directly measure spin diffusion lengths $\ell_{sf}^N$ for N = Cu(2.1 at.%Ge) and Ag(3.6 at.%Sn) alloys. We find $\ell_{sf}^{Cu(2\%Ge)} = 117^{+10}_{-6}$ nm and $\ell_{sf}^{Ag(4\%Sn)} = 39 \pm 3$ nm. The good agreement of this $\ell_{sf}^{Cu(2\%Ge)}$ with the value $\ell_{sf}^{Cu(2\%Ge)} = 121 \pm 10$ nm derived from an independent spin-orbit cross-section measurement for Ge in Cu, quantitatively validates the use of Valet-Fert theory for CPP-MR data analysis to layer thicknesses several times larger than had been done before. From the value of $\ell_{sf}^{Ag(4\%Sn)}$, we predict the ESR spin-orbit cross-section for Sn impurities in Ag.

The spin-diffusion lengths, $\ell_{sf}^F$ and $\ell_{sf}^N$, are the distances over which electrons flip their spins as they diffuse through ferromagnetic (F) or non-magnetic (N) metals. These lengths are fundamental parameters of F/N multilayers that play increasingly important roles in various magnetic devices. $\ell_{sf}^F$ and $\ell_{sf}^N$ can be measured by several techniques, including weak localization, current-perpendicular-to-plane (CPP) magnetoresistance (MR), and lateral non-local (LNL) resistance measurements [1-9]. Published derivations of $\ell_{sf}^F$ and $\ell_{sf}^N$ from CPP-MR or LNL data are based on free-electron models in which $\ell_{sf}^F$ and $\ell_{sf}^N$ are the only lengths, aside from the thicknesses or breadths of the F and N layers themselves. The applicability of such models was questioned on the basis of real Fermi surface calculations that yield layer-thickness-dependent interface resistances that scale exponentially with the mean-free-path—mean-free-path effects [10]. In fact, it was proposed that derivations of values of $\ell_{sf}^N$ for Fe$_{50}$Mn$_{50}$, V, Nb, and W, using the same measuring technique as in the present Letter, might be flawed, with the observed behaviors of the data being due instead to mean-free-path effects [11]. This issue was addressed in detail in Appendix C of ref. [1] for dilute alloys with values of $\ell_{sf}^N \leq 22.4$ nm. For these alloys, quantitative CPP-MR analyses were used to argue that mean-free-path effects are small. Alloys were used, because comparison values of $\ell_{sf}^N$ could be calculated completely independently using electron-spin-resonance (ESR) measurements of spin-orbit cross-sections [12] for the given impurity in the host metals Cu or Ag. In nominally 'pure' N-metals, in contrast, there is no independent way to calculate $\ell_{sf}^N$ for a given sputtered or evaporated sample, because a significant portion of the total scattering is produced by unknown concentrations of unknown residual impurities. Indeed, experimental values of $\ell_{sf}^N$ in such metals usually scatter widely, even for samples with similar residual resistivities [1]. Thus, while there is wide use of the Valet-Fert (VF) [13] and related models, and widespread belief in the reliability of the values of $\ell_{sf}^N$ that they produce, we contend that there is no prior proof of quantitative reliability beyond $\ell_{sf}^N = 22.4$ nm. Extending such a proof to $\ell_{sf}^N > 100$ nm is one of the values of the present study.

Prior CPP-MR studies of $\ell_{sf}^N$ in dilute alloys used two different measuring methods. Method I used multilayers of Co and the alloy of interest [5]. Its values of $\ell_{sf}^N$ for alloys with strong spin-flipping impurities, Cu(Pt), Cu(Ni), and Ag(Pt), agreed well with independent calculations using ESR spin-orbit scattering cross-sections [1]. For alloys with weaker spin-flipping impurities, Cu(4%Ge) and Ag(4%Sn), method I, which cannot distinguish too long $\ell_{sf}^N$ from $\ell_{sf}^N = \infty$, gave only lower bounds on $\ell_{sf}^N$. Method II used exchange-biased spin-valves (EBSVs) [6]. In the two cases where the same alloy was measured with both methods, Cu(6%Pt) and Cu(22.7%Ni), similar results were found [1].

Because it has no intrinsic limit on the length of $\ell_{sf}^N$ it can measure, in the present study, we use method II to extend the alloy test to $\ell_{sf}^N > 100$ nm. For such a test, we needed the alloy pair to satisfy three conditions: (a) large resistivity/atomic % impurity, so that a small impurity concentration will dominate the scattering; (b) small spin-orbit scattering, so that $\ell_{sf}^N$ will be long; and (c) an ESR value for the spin-orbit cross-section, letting the CPP-MR $\ell_{sf}^N$ be compared with an independently determined value. Both Ge in Cu and Sn in Ag give strong elastic scattering to satisfy (a) (Cu and Ag each have only one conduction electron, while Ge and Sn contribute four), but only weak spin-flipping to satisfy (b) (i.e., weak spin-orbit interactions, since Ge and Cu, and Sn and Ag, are in the same rows of the periodic table with the Z's of Ge and Sn only 3 larger than those of Cu and Ag). However, we have found an ESR spin-orbit cross-section only for Ge in Cu. A Sn content in Ag of 4% lets us directly compare our value of $\ell_{sf}^{Ag(4\%Sn)}$ with the lower bound measured by method I. A Ge content in Cu reduced to 2% gives an expected $\ell_{sf}^{Cu(2\%Ge)} > 100$ nm. Agreement between CPP-MR and ESR values of $\ell_{sf}^{Cu(2\%Ge)}$ would validate CPP-MR method II, and extension of Valet-Fert theory, to values of $\ell_{sf}^N$ longer than 100 nm, comparable to lengths measured with lateral spin-valves [1,3,4].

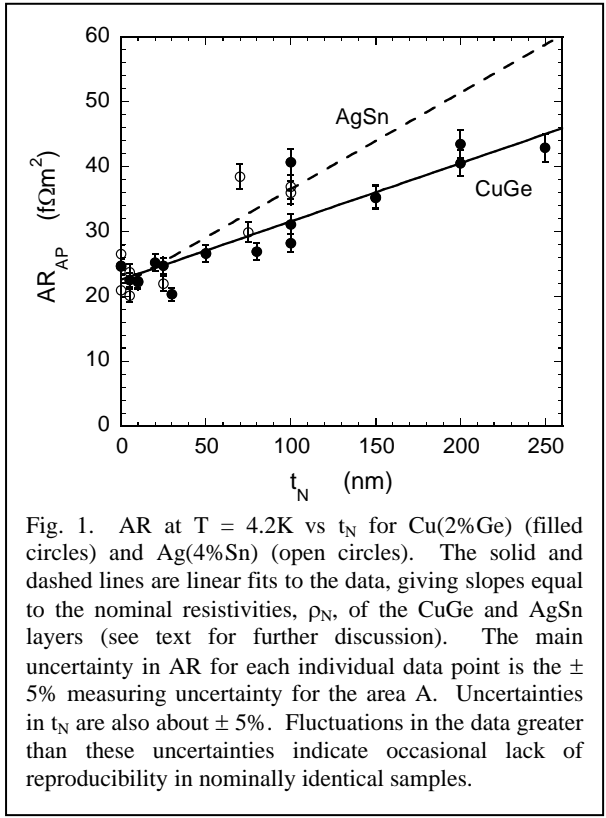

Fig. 1. AR at T = 4.2K vs $t_N$ for Cu(2%Ge) (filled circles) and Ag(4%Sn) (open circles). The solid and dashed lines are linear fits to the data, giving slopes equal to the nominal resistivities, $\rho_N$, of the CuGe and AgSn layers (see text for further discussion). The main uncertainty in AR for each individual data point is the ± 5% measuring uncertainty for the area A. Uncertainties in $t_N$ are also about ± 5%. Fluctuations in the data greater than these uncertainties indicate occasional lack of reproducibility in nominally identical samples.

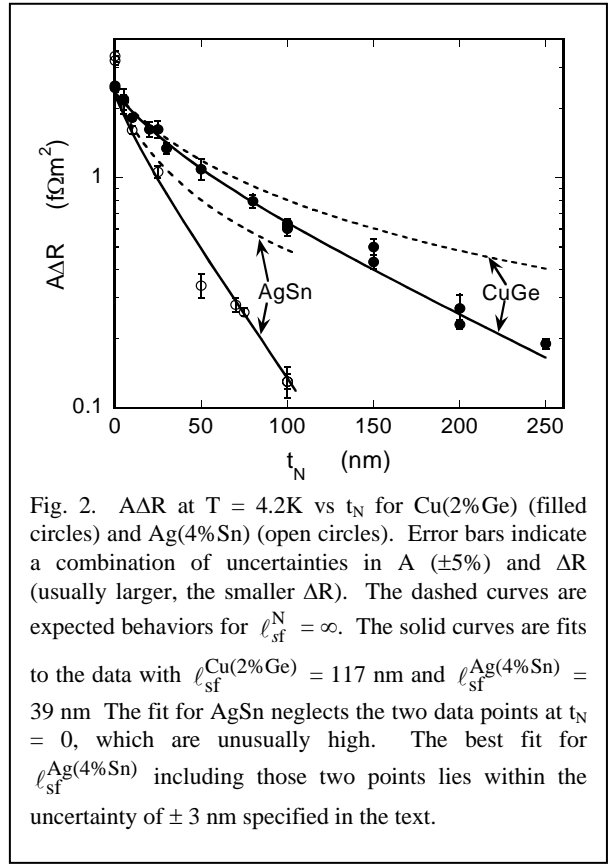

Fig. 2. A$\Delta$R at T = 4.2K vs $t_N$ for Cu(2%Ge) (filled circles) and Ag(4%Sn) (open circles). Error bars indicate a combination of uncertainties in A (±5%) and $\Delta$R (usually larger, the smaller $\Delta$R). The dashed curves are expected behaviors for $\ell_{sf}^N = \infty$. The solid curves are fits to the data with $\ell_{sf}^{Cu(2\%Ge)} = 117$ nm and $\ell_{sf}^{Ag(4\%Sn)} = 39$ nm  The fit for AgSn neglects the two data points at $t_N = 0$, which are unusually high. The best fit for $\ell_{sf}^{Ag(4\%Sn)}$ including those two points lies within the uncertainty of ± 3 nm specified in the text.

Following [6-8], our samples were EBSVs based upon the ferromagnet Py = $Ni_{1-x}Fe_x$ with x ~ 0.2, and the antiferromagnet, FeMn. The EBSVs had the form Nb(100)/Cu(5)/FeMn(8)/Py(24)/Cu(5)/N($t_N$)/Cu(5)/Py(24)/Cu(5)/Nb(100), where all thicknesses are in nm, and N is the non-magnetic alloy of interest, which has adjustable thickness $t_N$. Here the main multilayer, extending from the FeMn through the second Py layer, is sandwiched between two 100 nm thick, 1.1 mm wide, Nb cross-strips. At the measuring temperature of 4.2K, the Nb strips are superconducting, giving a uniform current flow through the multilayer of interest [14,15]. The Py layer adjacent to the FeMn is exchange-bias pinned to the FeMn by heating each sample to 453 K, applying a magnetic field H = 180 Oe in the plane of the layers, and then cooling the sample to room temperature in the presence of the field. The other Py layer is left free to reverse at lower applied field H (~ 20 Oe), allowing the sample to be toggled between the anti-parallel (AP) orientation of the two Py moments with high resistance, $R_{AP}$, and the parallel (P) orientation with low resistance, $R_P$.

The alloy sputtering targets were nominally N = Cu(2%Ge) and Ag(4%Sn). Impurity concentrations of 2.1 ± 0.3 at. % Ge and 3.6 ± 0.5 at. % Sn were estimated from the residual resistivities calculated below, using the values of resistivity per atomic percent impurity given in [16], after subtracting residual resistivities of 5 n$\Omega$m for our sputtered Cu [15] and 10 n$\Omega$m for our Ag [14]. Independent electron energy dispersive x-ray (EDS) measurements on the films were consistent with the assigned impurity contents to within mutual uncertainties.

The quantities of experimental interest are the specific resistances $AR_{AP}$, $AR_P$, and especially their difference

$$A\Delta R = AR_{AP} - AR_P, \qquad (1)$$

where A is the area of overlap of the two crossed Nb strips, through which the CPP current flows into and out of the sample EBSV. In the present experiments, all quantities in the EBSV are held fixed except $t_N$ of the inserted alloy. Fig. 1 shows plots of $AR_{AP}$ vs $t_N$ for both Cu(2%Ge) and Ag(4%Sn). Fig. 2 shows plots of $A\Delta R$ vs $t_N$ for the same two alloys.

We start with Fig. 1. Since increasing $t_N$ just adds an additional amount of the alloy N to the EBSV, we expect $AR_{AP}$ to grow closely linearly with $t_N$, and the slope of a plot of $AR_{AP}$ vs $t_N$ to give an estimate of the alloy resistivity $\rho_N$. Such plots are shown in Fig. 1 for both CuGe and AgSn. The best fit slopes are $\rho_{Cu(2\%Ge)}$ = 90 ± 10 nΩm and $\rho_{Ag(4\%Sn)}$ = 150 ± 30 nΩm. Independent van der Pauw [17] measurements of separately sputtered 200 nm thick films gave $\rho_{Cu(2\%Ge)}$ = 80 ± 4 nΩm and $\rho_{Ag(4\%Sn)}$ = 170 ± 25 nΩm. Combining these values, taking account of the specified uncertainties, gives best estimates of ρ used in our subsequent analyses of $\rho_{Cu(2\%Ge)} = 81^{+6}_{-2}$ nΩm and $\rho_{Ag(4\%Sn)} = 160 ± 20$ nΩm.

Analysis of the decreasing values of $A\Delta R$ in Fig. 2 is more complex, as described in detail in [6]. Here we focus on the two primary factors behind that decrease. The one of main interest is the increase in total spin-flipping within the N-insert as $t_N$ increases. This increase should cause $A\Delta R$ to decrease as $\exp(-t_N/\ell_{sf}^N)$. If this effect were the only one, a plot of ln ($A\Delta R$) vs $t_N$ would yield a straight line, the slope of which would give $\ell_{sf}^N$. In practice, $A\Delta R$ also decreases with increasing $t_N$ because of the extra specific resistance ($\rho_N t_N$) that is added to the EBSV [1,6,13,15]. To illustrate this latter effect, we show as dashed curves in Fig. 2 the estimated decreases of $A\Delta R$ with $t_N$ for assumed $\ell_{sf}^N = \infty$--i.e., assuming no decrease in $A\Delta R$ due to finite $\ell_{sf}^N$. We see that the experimental data fall well below these dashed curves. To analyse the data, and derive values of $\ell_{sf}^N$, we use equations given in [6], based upon the VF model of the CPP-MR [13]. We evaluate these equations numerically, using the parameters given above plus those in [6,8] for the general Py-based EBSV. For each alloy, the only unknown parameter in the analysis is $\ell_{sf}^N$. The best fits are the solid curves, which give the values $\ell_{sf}^{Cu(2\%Ge)} = 117^{+10}_{-6}$ nm and $\ell_{sf}^{Ag(4\%Sn)} = 39 ± 3$ nm. Here we specify twice the standard deviation, with the CuGe uncertainty biased toward the high side due to the uncertainty in its residual resistivity. This value for Ag(4%Sn) is consistent with the previous lower bound of $\ell_{sf}^{Ag(4\%Sn)} \geq$ 26 nm [1,5], and this value for Cu(2%Ge) is consistent with rescaling to $\ell_{sf}^{Cu(2\%Ge)} \geq$ 100 nm of the previous lower bound for Cu(4%Ge) of $\ell_{sf}^{Cu(4\%Ge)} \geq 50$ nm [1,9].

Finally, we compare the value of $\ell_{sf}^{Cu(2\%Ge)} = 117^{+10}_{-6}$ nm with what we expect from electron spin-resonance (ESR) measurements of the spin-orbit cross-section of Ge in Cu [12]. We use the VF equation [13]

$$\ell_{sf}^N = \sqrt{(\lambda_N \lambda_{sf})/6}, \qquad (2)$$

relating $\ell_{sf}^N$ to the elastic mean-free-path, $\lambda_N$, and the spin-flip mean-free-path, $\lambda_{sf}$. For those interested in the equivalent relaxation times, within the same model as described above, $\lambda_N = v_F \tau_N$ and $\lambda_{sf} = v_F \tau_{sf}$ relate the elastic mean-free-path to the momentum relaxation time, and the spin-flip mean-free-path to the spin-flip relaxation time, via the Fermi velocity $v_F$. But we determine both $\lambda_N$ and $\lambda_{sf}$ without explicit reference to relaxation times. We calculate $\lambda_N = 0.66$ fΩm²/$\rho_N$ = 8.2 nm using the numerator for Cu [16]. We calculate $\lambda_{sf} = 10.8 \times 10^3$ nm from the ESR spin-orbit cross-section for Cu(Ge) from [12]. Our resulting spin-diffusion length, $\ell_{sf}^{Cu(2\%Ge)} = 121 ± 10$ nm, overlaps with our newly measured value to within experimental uncertainties. This agreement provides support for the spin-diffusion length $\ell_{sf}^{Cu(5\%Ge)} = 55 ± 5$ nm used in [18] for a Cu(5%Ge) alloy derived from the same ESR spin-orbit cross-section. We do not know of any ESR measurement of the spin-orbit cross section for Ag(Sn). In this case, we can invert the process we just used for Cu(Ge), and input our measured value of $\ell_{sf}^{Ag(4\%Sn)}$, to give what should be a reliable 'prediction' of the spin-orbit cross-section for Sn in Ag. Inverting Eq. (2) and using $\lambda_N = 0.84$ fΩm²/$\rho_N$ = 5.2 nm[15], we predict $\sigma_{Ag(Sn)} = (3 ± 1) \times 10^{-18}$ cm².

To summarize, we have used the CPP-MR technique with EBSVs of [6] to measure the spin-diffusion lengths, $\ell_{sf}^N$, for sputtered, dilute Cu(2%Ge) and Ag(4%Sn) alloys. We find values $\ell_{sf}^{Cu(2\%Ge)} = 117^{+10}_{-6}$ nm and $\ell_{sf}^{Ag(4\%Sn)} = 39 ± 3$ nm. Both values are consistent with the lower bounds previously found using a different technique [1,5], and the value for CuGe is consistent with that calculated from the independently measured spin-orbit cross-section for Ge in Cu [12]. In addition to providing specific values of the spin-diffusion lengths in dilute CuGe and AgSn alloys, by showing that the technique of ref. [6] can be reliably extended to a spin-diffusion length greater than 100 nm, these measurements provide new, quantitative support for the use of Valet-Fert theory to analyze CPP-MR measurements at such large thicknesses, and of equivalent analyses for Lateral-non-local (LNL) measurements. Lastly, the good agreement, both here and in ref. [1], between measured values of $\ell_{sf}^{N}$ for various Cu- and Ag-based alloys, and those calculated from ESR measurements, means that our new value of $\ell_{sf}^{Ag(4\%Sn)}$ should let us reliably predict the spin-orbit cross-section for Ag(Sn) which, to our knowledge, has not been measured. We predict $\sigma_{Ag(Sn)} = (3 \pm 1) \times 10^{-18}$ cm$^2$.

Acknowledgments: We thank N.O. Birge for helpful suggestions. This work was supported by US NSF Grants DMR-05-01013 and PHY-024-3709.

References.
1. J. Bass and W.P. Pratt Jr., J. Phys. Cond. Matt. **19**, 183201 (2007).
2. G. Bergmann, Z. Phys. **B48**, 5 (1982).
3. M. Johnson and R.H. Silsbee, Phys. Rev. Lett. **55**, 1790 (1985)
4. F.J. Jedema, A.T. Filip, and B. J. van Wees, Nature **410**, 345 (2001).
5. Q. Yang, P. Holody, S.-F. Lee, L. L. Henry, R. Loloee, P.A. Schroeder, W.P.Pratt Jr.,and J. Bass, Phys. Rev. Lett. **72**, 3274 (1994).
6. W. Park, D. Baxter, S. Steenwyk, I. Moraru, W.P .Pratt Jr., and J. Bass, Phys. Rev. **B62**, 1178 (2000).
7. S.D. Steenwyk, S.Y. Hsu, R. Loloee, J. Bass, and W.P. Pratt Jr., J. Magn. Magn. Mat. **170**, L1 (1997).
8. L. Vila, W. Park, J.A. Caballero, D. Bozec, R. Loloee, W.P. Pratt Jr., and J. Bass, J. Appl. Phys. **87**, 8610 (2000).
9. J. Bass, P.A. Schroeder, W.P. Pratt Jr., S.F. Lee, Q. Yang, P. Holody, L.L. Henry and R. Loloee, Mater. Sci. Eng. B **31** 77 (1995).
10. E.Y. Tsymbal and D.G. Pettifor, Solid State Physics Series, **56**, 113 (2001).
11. Ref. [10], page 233.
12. P. Monod and S. Schultz, J. Physique **43**, 393 (1982).
13. T. Valet and A. Fert, Phys. Rev. **B48**, 7099 (1993).
14. S.-F. Lee, Q. Yang, P. Holody, R. Loloee, J.H. Hetherington, S. Mahmood, B. Ikegami, K. Vigen, L.L. Henry, P.A. Schroeder, W.P. Pratt Jr., and J. Bass, Phys.Rev. **B52**, 15426 (1995).
15. J. Bass and W.P. Pratt Jr., J. Magn. Magn. Mat. **200**, 274 (1999).
16. J. Bass, Landolt-Bornstein Tables, New Series, **III/15a**, Springer-Verlag Publ., Pgs. 1-288 (1982).
17. L. van der Pauw, Philips Res. Rep. **13**, 1 (1958).
18. N. Theodoropoulou, A. Sharma, W.P. Pratt Jr., J. Bass, M.D. Stiles, J. Xiao, Phys. Rev. **B76**, 220408(R) (2007).